\documentstyle [titlepage,12pt,seceqn]{article}
\begin{document}
\pagestyle{plain}
\baselineskip=0.62cm
\begin{flushright}
{Crete$-$96$-$10}\\
{hep$-$th$/$9601172}
\end{flushright}
\vspace{1.5cm}
\begin{center}
{\large 
VORTEX PAIRS IN CHARGED FLUIDS
}
\end{center}
\vspace {1.2cm}

\centerline{ {\large G.N. Stratopoulos  \footnote{
email: stratos@physics.uch.gr} } }
\vskip 0.4cm
\centerline {and}
\vskip 0.4cm 
\centerline{ {\large T.N. Tomaras  \footnote{
email: tomaras@physics.uch.gr  } } }  
\vskip 0.4cm 

\centerline {\it {Department of Physics, University of Crete 
and Research Center of Crete,}}
\centerline {\it {P.O.Box 2208, 710 03 Heraklion, Crete; Greece }} 
\vskip 1.5cm
\centerline{\large\bf abstract}

The motion of a vortex-(anti)vortex pair
is studied numerically in the framework of a dynamical 
Ginzburg-Landau
model, relevant to the description of a 
superconductor or of an idealized bosonic plasma.
It is shown that up to a fine "cyclotron" internal motion, 
also studied in
detail, 
two vortices brought together, rotate 
around each other,
while a vortex and an antivortex
move in formation parallel to each other.
The velocities of the vortices in both cases are measured to be
in remarkable agreement with
recent theoretical predictions, down to intervortex distances
as small as their characteristic diameter.

\newpage

\section{Introduction}

The dynamics of flux-vortices in ordinary \cite{rHuebener} 
as well as 
high$-T_C$ superconducting films \cite{rVinokur}
under the influence of a variety of external probes and for a wide  
range of temperatures,
has been an area of vigorous experimental and theoretical 
research during the past few decades. Their pinning by and 
depinning from
the impurities of the lattice, the role of their motion in energy 
dissipation,
the statistical mechanics of a vortex lattice or of a vortex 
glass have
received considerable attention. But still no unambiguous derivation
of the equation of motion of an 
isolated vortex is available \cite{rMagnus}. Correspondingly, 
issues such as
the existence and the origin of the Magnus force on a vortex, 
or the interpretation of the so called "opposite sign Hall 
effect" \cite{rOpposite} reported in all types of 
superconducting materials as they are cooled below their critical
temperature, have not yet been resolved.

In a recent publication \cite{rPTb} 
we proposed to study the motion of these vortices
in the framework of a phenomenological
effective field theoretic model \cite{rFeynman}, 
which combines the successes of the static Ginzburg-Landau 
theory of
superconductivity with those of the Gross-Pitaevskii description 
of the dynamics
of a neutral bosonic quantum liquid.
It should be pointed out, that several attempts
to actually derive the effective action for the superconductor 
from the fundamental theory 
have been reported recently \cite{rATZ}.
Their results are quite encouraging for the importance  
of the above model.

It has been argued on general grounds that the model 
should support the existence 
of absolutely stable flux-vortices \cite{rPTb}, similar in nature
to the previously known Abrikosov vortices \cite{rAbrikosov} 
of the static Ginzburg-Landau theory.
The solutions were found numerically and 
their static properties as well as the properties of pairs
of them have been studied in detail \cite{rST}. Although they
 carry zero
total electric charge, they have non-vanishing charge density
and electric field and thus differ significantly in the 
details from
all the previously known ones. 
Even more importantly, in the framework of the present 
model and for reasonably
rigid vortices or equivalently under 
external conditions not too violent 
to destroy the sense of a localized 
soliton, it was possible to derive 
analytically the equation
of motion of the vortex as a whole, 
under the influence of any kind of 
external force \cite{rPTb}, \cite{rST}.
The equation is rather unique, since it is equivalent to the
statement of momentum conservation.
It is then straightforward to obtain an approximate 
description of the motion
of a vortex-(anti)vortex pair.
For the resulting theoretical picture though, to be useful in 
realistic applications it is important to have some
control over the magnitude of the deviations from it.
The purpose of the present paper is twofold: First, 
to verify numerically the analytical predictions, and 
second, to explore
the range of applicability of the assumptions 
behind their derivation.
 
The plan of the presentation is as follows: 
Section 2 contains a general introduction to the model as well as
the study of the spectrum of small fluctuations around the vacuum. 
Its relevance
to the physics of a superconductor is also commented upon.
A brief review of the theoretical analysis of the vortex dynamics
is the content of section 3. 
The correct definition of the conserved
quantities in the non-trivial topological 
sectors of the theory and
their physical meaning is presented here, together with the main
predictions about the motion of the vortices in a vortex pair or in
a vortex-antivortex system. 
The results of the numerical simulations 
of the motion of the vortices
in these systems are presented in detail in section 4. 
Apart from the
verification of the qualitative picture, 
one also obtains a remarkable
agreement with the a priori approximate theoretical
formulas even for vortices overlapping considerably.
The Hall motion observed here is a common characteristic shared by
many systems discussed in the closing section.

\vspace{0.2cm}
\section{ The model - General properties }

The model describes the dynamics of a non-relativistic 
charged scalar field $\Psi$, minimally coupled to
the electromagnetic potential $(A_0, A_i)$.
The Lagrangian density is:
\begin{equation} 
{\cal L}\;=\; {i \gamma\over 2} 
[ \Psi \sp\ast {\cal D}_t\Psi \;-\; c.c. ]
 \;+\;\gamma q \Psi_0\sp2 A_0 
 -\; {\gamma\sp2\over 2m} \bigl| {\cal D}_i\Psi |^2  \\
+ {1\over 8 \pi} (\epsilon {\bf E}\sp2 - {\bf B}\sp2 )
\;-\; V(|\Psi|)
\label{lagrangian} 
\end{equation} 
\noindent
The magnetic and the electric field are respectively 
$ {\bf B} = \nabla \times {\bf A} $ and 
$ E_i = -{1\over c} \partial_t A_i - \partial_i A_0\;$,
while $\; {\cal D}_t\Psi=
(\partial_t+i q A_0)\Psi \;$,$\; {\cal D}_i\Psi=
(\partial_i-i {q\over c} A_i)\Psi$. 
A quartic phenomenological potential 
$V(|\Psi|)= 
{1\over 8} \;g\; (\Psi \Psi\sp\ast-\Psi_0\sp2)\sp2 $
may or may not be present, depending on the physical system of
interest.
$\;\gamma,\; m,\; \epsilon,\; g$ and $q$ are 
parameters, c is the speed of  
light and the spatial indices $i,\;j$ range from 1 to 
the dimensionality
of space. For simplicity 
we did not include an arbitrary parameter
in front of the term ${\bf B}^2$. Up to this inessential for our
purposes restriction,
the model is the most general theory possessing 
translational, rotational and gauge symmetry.
To make the model consistent we have included a  
background (positive-ion) charge
density $\; q \Psi_0\sp2 $  to neutralize the system. 
We work in the limit 
where ions are very heavy and the background is taken nondynamical.

With the identification of the field $\Psi$ 
as the condensate wave-function
of the Cooper pairs, and correspondingly with $\gamma\,=\,\hbar$, 
$m\,=\,2 m_e$ and $q\,=\,{2 e}$/${\hbar}$, 
(\ref{lagrangian}) becomes a 
reasonable phenomenological model of a superconductor. It offers 
a natural explanation of the Meissner (see (2.15) below) 
and the Josephson effects, and it
predicts the correct value of the quantum 
$\phi_0 \,=\, 2.09 \times 10^{-7}$
Gauss-c${\rm m}^2$ of the vortex magnetic flux 
\cite{rFeynman}. Furthermore, one may for simplicity set 
$\kappa\,=\,0$ and use the typical 
values of the background charge density and of the energy gap of a 
superconductor to determine the remaining parameters $\Psi_0$ and 
$\epsilon$ of the model. The values of the penetration
depth and the coherence length are then fixed 
and nicely fall within the
range characteristic of an ordinary type-II
superconductor \cite{rS}. 
For instance, the typical time and length, 
defined via (\ref{rescalings})
below, are of the order of $10^{-10}-10^{-8}$ sec and 
$50-500$ Angstrom, respectively.

\noindent
We switch to dimensionless fields and coordinates by the rescalings 
\begin{displaymath}   
x_i\to {\sqrt{m} c \over \sqrt{4 \pi} 
\Psi_0 q \gamma} \;\tilde x_i \hskip 1.4cm 
 t\to {m\sp2 c\sp2 \over 4\pi \Psi_0\sp2 q\sp2 
 \gamma\sp3} \; \tilde t   
\end{displaymath}
\begin{equation} 
\Psi\to \Psi_0   \tilde \Psi \hskip 1.4cm 
A_0\to {4\pi \Psi_0\sp2 q \gamma\sp3 \over 
m\sp2 c\sp2} \;\tilde A_0\hskip 1.4cm 
A_i\to {\sqrt{4 \pi} \Psi_0 \gamma \over \sqrt{m} } \; \tilde A_i 
\label{rescalings} 
\end{equation} 
\noindent
and write the Lagrangian in the form:
\begin{equation} 
{\cal L}\;=\;{1\over 2}
( \tilde \Psi\sp\ast (i\tilde \partial_t - \tilde A_0)\tilde \Psi
 \;+\; c.c.)\;+\;\tilde A_0  -\; {1\over 2} \;
|\tilde D_i\tilde \Psi|^2 + {1\over 2} ({1\over \beta} {\bf 
\tilde E}\sp2 - \tilde  {\bf B}\sp2 )
\;-\; {1\over 8} \kappa\sp2 (\tilde \Psi \tilde \Psi\sp\ast-1)\sp2
\label{dimensionlesslagrangian} 
\end{equation} 
\noindent
with $ \tilde {\bf B} = \tilde {\bf \nabla} \times 
\tilde {\bf A} $,
$\tilde  E_i = -\tilde \partial_t \tilde A_i -
\tilde \partial_i \tilde A_0 $ and 
$\tilde D_i=\tilde \partial_i-i \tilde A_i$ .  
The two remaining parameters, 
the quartic self-coupling $\kappa^2$, and the coupling
$\beta$ of the scalar field to the 
electrostatic potential, are defined by 
\begin{equation}  
\kappa\sp2 = {g m\sp2 c\sp2\over 4\pi q\sp2 \gamma\sp4}
\hskip 1.5cm \beta= {m\sp3 c\sp4 \over 
{4 \pi \epsilon q\sp2 \gamma\sp4  \Psi_0\sp2} }
\label{parameters} 
\end{equation} 
\noindent
Notice that we have changed the name $\lambda$ 
we used in \cite{rST} for the second
parameter to $\beta$, in order to avoid confusion with the standard
notation of the penetration depth
of a superconductor. 
Furthermore, to simplify the formulas 
we will drop the tilde henceforth. 

The action $S$, the integral of $\cal L$ over space and time,
is invariant under the 
gauge transformation
\begin{eqnarray}   
{\Psi}^{\prime}\;=\;exp{(i \Lambda ({\bf x}, t))} \Psi \nonumber \\
{A_i}^{\prime}\;=\;A_i+\partial_i 
\Lambda  \label{gaugetransformation} \\
{A_0}^{\prime}\;=\;A_0- \partial_t \Lambda \nonumber 
\end{eqnarray}
\noindent
for arbitrary function $\Lambda ({\bf x}, t)$,
and its extremization with respect to $A_0$ 
leads to the Gauss constraint: 
\begin{equation}
{1\over \beta} \;\partial_i E_i\;=\; \Psi \Psi\sp\ast-1
\label{gaussconstraint} 
\end{equation}   
\noindent
We shall only be interested in configurations with vanishing total
electric charge.
The equations of motion derived by varying $S$ 
(under this constraint) 
with respect to $ \Psi \sp\ast $ and $ A_i $ read
\begin{displaymath} 
i \dot \Psi=-{1\over 2} {{\bf D}\sp
2} \Psi + A_0 \Psi+{1\over 4}
{\kappa}^2 (\Psi\sp\ast  \Psi-1)\Psi   
\end{displaymath}
\begin{equation} 
{1\over \beta}\dot {\bf E} = \nabla\times {\bf B} - {\bf J} 
\label{equationsofmotion} 
\end{equation} 
\noindent
with the current 
${\bf J} = [\Psi\sp\ast {\bf D} \Psi - c.c.] / {2 i}$.
The energy $W$ of an arbitrary configuration of the system 
is the spatial
integral of the energy density $w$, which 
is the sum $w=w_d+w_b+w_e+w_v$
of the four positive definite terms:
\begin{equation}   
w_d= {1\over 2}\; |D_i \Psi|\sp2  
\hskip 1.cm 
w_b={1\over 2}\;{{\bf B}^2} 
\hskip 1.cm 
w_e={1\over 2 \beta}\; {{\bf E}\sp2} 
\hskip 1.cm
w_v={1\over 8 }  \kappa\sp2 \;(\Psi \Psi\sp\ast-1)\sp2
\label{energy} 
\end{equation} 

\vspace{0.3cm}

{\it  The vacuum }

Equations  (\ref{gaussconstraint}) and (\ref{equationsofmotion}) 
admit the one parameter 
family of equivalent vacuum solutions
\begin{equation} 
\Psi= e^{i \alpha}  \hskip 1.2cm A_i=0 \hskip 1.2cm A_0=0 
\label{vacuum} 
\end{equation} 
\noindent
parametrized by the constant angular parameter $\alpha$.

To study the spectrum of small fluctuations 
around the vacuum solution
we choose the one with $\alpha=0$ and 
the Coulomb condition ${\bf \nabla}\cdot {\bf A} = 0$ to remove the 
gauge arbitrariness of the model.
We then parametrize the generic deviation of 
$\Psi$ from its vacuum value $\Psi=1$ by:
\begin{equation}    
\Psi \;=\; (1 + \Phi) e\sp  { i \Theta }   
\label{fluctuations} 
\end{equation} 
\noindent
For the discussion of small fluctuations 
the magnitudes of $\Phi$, 
$\Theta$, $A_0$ and $A_1$ will all be taken much smaller than one.
\noindent
Keeping only up to quadratic terms in these small fields,
the Lagrangian becomes:
\begin{equation}	
{\cal L}\;=\;
  -\partial_t\Theta (1+2\Phi) - 2\Phi A_0 - {1\over 2} 
  (\partial_i\Theta^2 +  \partial_i\Phi^2 + A_i^2)
  + {1\over 2\beta} (\partial_i A_0^2 + \partial_t A_i^2 )
  - {1\over 2} {\bf B}\sp2 - {1\over 2} \kappa\sp2 \Phi^2
\label{S3}   
\end{equation} 
\noindent
and the equations of motion read
\begin{equation} 
\partial_t\Phi + \partial_i^2\Theta \;=\; 0
\label{S4}  
\end{equation} 
\begin{equation}  
\partial_t\Theta \;=\; {1\over 2} \partial_i^2\Phi - A_0  -
{1\over 2} \kappa\sp2 \Phi
\label{S5}  
\end{equation} 
\begin{equation}
{1\over \beta} \partial_i^2 A_0 + 2\Phi \;=\; 0
\label{S6}  
\end{equation} 
\begin{equation}  
{1\over \beta} \partial_t^2 A_i  \;=\;  \partial_k^2 A_i - A_i
\label{S7} 
\end{equation} 
\noindent
The field $A_i$ decouples at this level.
Acting with the time derivative on (\ref{S4}), with the Laplacian
on (\ref{S5}) and using (\ref{S6}), one obtains the following 
equation of motion of $\Phi$:
\begin{equation}    
\partial_t^2\Phi \;=\;  {1\over 4}  \kappa^2 \partial_i^2 \Phi
- \beta \Phi - {1 \over 4} \partial_i^4 \Phi
\label{S8} 
\end{equation} 
\noindent
The plane-waves  
$\, A_i = A_i^0 \; e^{-i ({\omega_A} t - 
{\bf k}\cdot {\bf x})} \, $ 
and
$\, \Phi = \Phi^0 \;e^{- i ({\omega_\Phi} t - 
{\bf k}\cdot {\bf x} ) } \, $ 
obeying the dispersion relations
\begin{equation}  
\omega_\Phi^2 \;=\;  \beta +  {1 \over 4} \kappa^2  |{\bf k}|^2 + 
{1 \over 4}  |{\bf k}|^4 
\label{dispersion1} 
\end{equation} 
\begin{equation}  
\omega_A^2 \;=\; \beta  (1 + |{\bf k}|^2)
\label{dispersion2} 
\end{equation} 
\noindent
form a complete set of solutions of the equations of 
motion of $\Phi$ and $A_i$
above. Notice that both spectra $\omega_\Phi({\bf k})$ and 
$\omega_A({\bf k})$
have an energy gap $G\,=\, \sqrt \beta $.   
Finally, the solutions for $\Theta$ and $A_0$ 
are obtained by solving
(\ref{S4}) and (\ref{S6}), respectively.

\section{ Vortex Dynamics }

The model
under study, with or without the potential term $V(|\Psi|)$ 
present, supports the existence of flux-vortex solutions.
They are infinitely long, localized in the transverse direction, 
smooth, cylindrically symmetric, 
z-independent configurations with finite
energy per unit length,
whose static properties together with the properties of
pairs of them, have been studied in detail \cite{rST}.
We wish to study their dynamics numerically
and to verify the approximate analytical 
predictions about their motion \cite{rPTb}, 
reviewed briefly in the present
section. We will ignore the z-dependent excitations 
of the string and consequently
the formalism reduces to purely
$2+1$ dimensional. Spatial indices 
will from now on take the values 1 and 2,
while the magnetic field will only have 
its third component non-vanishing and
will be simply denoted by $B$.
Correspondingly, we will be thinking of the vortices
as finite energy "particle-like" localized 
objects in two spatial dimensions.
Any finite energy configuration is characterized by
an absolutely conserved integer number $N$, which counts the 
number of times the phase of the scalar field at spatial infinity,
a function of the polar angle $\theta$, 
winds around the circle of vacua 
(\ref{vacuum}), as $\theta$ varies from
zero to $2 \pi$. It is evaluated for the given configuration by
integrating over space a topological density $\tau({\bf x})$. 
Among the various possibilities 
the most useful form of $\tau$ is, for our 
purposes, the manifestly gauge invariant expression:
\begin{equation} 
\tau\;=\;{1\over 2 \pi i}
\;[\epsilon_{kl} (D_k\Psi)\sp\ast (D_l\Psi)-i B 
(\Psi\sp\ast \Psi-1)]  
\label{topologicaldensity} 
\end{equation} 
\noindent
As it will be shown immediately, this quantity 
appears in the formulas for
the conserved momentum and angular momentum of the theory.

Indeed, it was pointed out in 
reference \cite{rPTb}, that the naive Noether
expressions for the linear and the angular momentum of the model
are ambiguous for any configuration 
with non-zero topological charge,
and that the correct formulas for these quantities are:
\begin{equation} 
P_k=\epsilon_{ki} \int d^2x\; (2 \pi\; x_i\; \tau + 
{1\over \beta}\;E_i B)
\label{momentum} 
\end{equation} 
\noindent
and
\begin{equation} 
l=-\int d^2x\; ( \pi {\bf x}^2 \tau + {1\over \beta}\;
{\bf x}\cdot {\bf E} B)
\label{angularmomentum} 
\end{equation} 
\noindent
respectively. 
They differ from the naive expressions by surface terms,
which are important in topologically non-trivial sectors.
The presence 
of the first and the second moments of the
topological density $\tau$ in the above formulas, 
inherits them with an entirely different
physical meaning. 
Let us consider the momentum. Rotate it by $90^0$ and divide by the
constant $2 \pi N$ to obtain the new conserved quantity:
\begin{equation}  
R_i\equiv -{1\over 2\pi N} \epsilon_{ij} P_j \label{guidingcenter} 
\end{equation} 
\noindent
The value of ${\bf R}$ for an isolated 
axially symmetric vortex solution
is exactly the center of the vortex, 
and a rigid displacement by ${\bf c}$ 
of any given configuration as a
whole changes ${\bf R}$ by ${\bf c}$. 
Thus, the natural interpretation
of ${\bf R}$ is
the "mean position" of a generic localized 
configuration and this
justifies the name
${\it guiding \; center}$ for it.
Similarly, the first term of the angular momentum is a measure of
the size of the localized configuration 
and not of its rotational motion. 

Finally, one should mention the fact that the two components of the
momentum, the generators of translations 
in the x and the y directions,
do not commute. Instead, their Poisson bracket is
\begin{equation}
\{P_1, P_2\}=2 \pi N             \label{momentumalgebra} 
\end{equation} 
\noindent
a property suggesting a deep algebraic similarity of the model 
under study to a 
system of electric charges moving in a plane and
in the presence of a perpendicular magnetic field \cite{rPTa}.
Up to a multiplicative constant, 
the topological charge plays the role 
of the external magnetic field of the analogue model.

In the absence of external forces ${\bf P}$ is constant and so is
$\bf R$. Thus, a free localized vortex whose 
mean position is given by ${\bf R}$
will be spontaneously pinned at its initial position. No 
free translational motion of a vortex is possible. 
Under the influence of an external force ${\bf F}$, the momentum
evolves according to Newton's law $dP_k/dt\,=\,F_k$ and this 
translates into the following equation of motion of the vortex:
\begin{equation} 
{d\over dt} R_k=-{1\over {2 \pi N}} \;\epsilon_{kl} \;F_l
\label{vortexvelocity} 
\end{equation} 
\noindent
i.e. the vortex moves with speed ${\bf |F|}/{2 \pi |N|}$ and at
$\pm\; 90^0$ relative to the force for positive 
and negative $N$, respectively.
We see that the vortices exhibit the Hall behaviour known from
the analogue electric charge system mentioned above. 
This is how the analogy of the canonical structures of the 
two systems is reflected in the dynamics.

Based on (\ref{vortexvelocity}), 
one may immediately conclude that 
the vortices of a vortex pair will
rotate around each other. 
A rough theoretical estimate of the corresponding angular velocity 
$\Omega$ is easily obtained, especially when 
both vortices carry the same number $N$ of flux quanta.
Define for each one of the two vortices 
its approximate guiding center 
by (\ref{guidingcenter}) and (\ref{momentum}) with the integral 
taken over the corresponding half-plane. 
For localized vortices separated
by a distance $d$, large compared to their characteristic size, 
this is a reasonable definition of their positions.
Let us further assume that each vortex behaves more or less 
like a rigid body 
living in a potential equal to the vortex-vortex
interaction energy $U_{\rm vv}(d)$. Then, the magnitude of $\Omega$ 
is given by: 
\begin{equation}
\Omega(d) = {1\over {\pi |N| d}} \big| U_{\rm vv}^\prime (d) \big| 
\label{Omega}
\end{equation}   
\noindent
and its direction is counterclockwise (clockwise) for 
$N \bf \Delta\cdot F$ positive (negative), respectively. 
The vector $\bf \Delta$ joins the center 
of the system to the approximate position of
any one of the two vortices, while ${\bf F}$ 
denotes the force acting on it. 
 
In the same spirit, a vortex-antivortex pair with $N(-N)$ flux
units respectively, is expected 
to move in formation in a 
direction perpendicular to the line connecting them,
and with a speed given by 
\begin{equation}
V(d)\,=\, {1\over {2 \pi |N|}} \big| 
U_{\rm v\bar v}^\prime (d) \big|
\label{speed}
\end{equation} 
\noindent
The direction of their motion coincides with that of 
${\bf r} \times {\hat {\bf z}}$, where 
${\bf r} \equiv {\bf R}_+ - {\bf R}_-$ the vector 
joining the negative
to the positive-flux vortex, and ${\hat {\bf z}}$ 
the unit vector out of
the plane.
Incidentally, one may check that, like in the 
relativistic Abelian-Higgs
model \cite{rC}, the vortex and the antivortex
attract at all distances, a fact used above in 
the determination of the
direction of motion of the pair.

Clearly, formulae (\ref{Omega}) and (\ref{speed}) 
should not a priori be trusted for very small separations
of the two solitons. 
Although for $d$ much larger than
their characteristic size the hypotheses behind their
derivation are physically sound, 
smaller $d$'s make such
approximations questionable.
It will be shown though in the next section, 
through a direct quantitative comparison
of these formulas with our numerical results, that 
(\ref{Omega}) and (\ref{speed}) are reliable and describe
quite accurately 
the vortex-(anti)vortex motion 
even at distances as small as their characteristic diameter.

\section { Numerical Results }

{\it (a)  Discretization}

We now turn to the numerical simulation
of the motion of a pair of vortices and of 
a vortex-antivortex system
due to their mutual interaction. 
We choose to discretize the system in a way
that preserves as much of the symmetry of the continuous theory as
possible. In particular, as explained in the Appendix,  
it is convenient 
to preserve the gauge invariance of the model. 
Otherwise one has difficulty
in imposing Gauss' local constraint, 
and this leads to integration instabilities.
But then, naive discretization of the model is not appropriate and
one has instead to use 
techniques developed in the study of lattice
gauge theories \cite{rMCarlo}.

Space is replaced by a two 
dimensional square lattice with lattice
spacing $a$. The scalar field 
is replaced by the 
variables $ \Psi_{i,j} $, all functions of time,  
which live on the vertices  of the lattice.
Similarly, the spatial components 
of the gauge field are represented by
$ A^1_{i,j}, A^2_{i,j} $ and live on the corresponding 
oriented link connecting $ (i,j) $ to $ (i+1,j) $ and to
$ (i, j+1) $, respectively.
At this stage time is left continuous 
and the electrostatic potential 
$ A^0_{i,j} $ lives on the vertices of the grid.
 
The lattice version of the covariant derivative is:
\begin{equation}
D_k\Psi_{i,j} \;=\; {1\over a}(U^k_{i,j} \Psi_{(i,j)+\hat{k}} - 
\Psi_{i,j})
\hskip .6cm {\rm where}  \hskip .6cm  U^k_{i,j} = 
{\rm exp}(-i a A^k_{i,j})
\label{covariantderivative}
\end{equation} 
\noindent
and the lattice action takes the form:
\begin{eqnarray} 
  S \;=\; 
  \int dt \; a^2 \; \sum_{i,j} \; \{ \; {1\over 2} 
  [\Psi^{\ast}_{i,j}
  ( i \partial_t -  A^0_{i,j}) \Psi_{i,j} \; + c.c.] \; + A^0_{i,j}
  \; - {1\over 2} |D_k \Psi_{i,j}|^2 \;+ {1\over 2 \beta} 
  {\bf E}^2_{i,j}
   \nonumber \\
  \;-\;  {1\over 8} \kappa\sp2 
  (\Psi_{i,j}\sp{\ast}  \Psi_{i,j} - 1)\sp2
 \;-\;{1\over  a^4} 
 [1 -cos(a ( A^1_{i,j}+A^2_{i+1,j}-A^1_{i,j+1}-A^2_{i,j} ))] \;\;\}
\label{latticeaction} 
\end{eqnarray}
\noindent
where $ E^k_{i,j} \;=\; - \partial_tA^k_{i,j} - {1\over a} 
(A^0_{(i,j)+\hat{k}}-A^0_{(i,j)})$. In the
$a \to 0$ limit the 
last term of (\ref{latticeaction})  
becomes the ${1 \over 2}B^2$ term of the continuous model.

The action (\ref{latticeaction}) is indeed  
invariant under the discrete version of the gauge
transformation (\ref{gaugetransformation})
\begin{eqnarray}
& \Psi_{i,j}\to {\rm exp} (i\Lambda_{i,j}) \Psi_{i,j} & \nonumber \\
& A^k_{i,j}\to A^k_{i,j} + {1 \over \alpha} 
( \Lambda_{(i,j)+\hat{k}} - 
\Lambda_{i,j})  & \label{latticegaugetransformation} \\
& A^0_{i,j}\to A^0_{i,j} - \partial_t \Lambda_{i,j} &  \nonumber 
\end{eqnarray}
\noindent
Correspondingly, the system obeys the discretized form of 
Gauss' law:
\begin{equation}
{1\over a} \sum_{k=1,2} \; \{\; (E^k_{(i,j)+\hat{k}}-E^k_{i,j}) 
\;\} \;=\; 
\beta (\Psi_{i,j}\sp{\ast} \Psi_{i,j} - 1)
\label{latticegauss} 
\end{equation} 
\noindent
where $ \hat{k} $ is the unit 
lattice step in the spatial direction $k$.
Variation of the lattice action (\ref{latticeaction}) 
with respect to $ \Psi_{i,j}\sp\ast $
and $ A^k_{i,j} $ leads to the equations of motion:
\begin{displaymath}
i \dot{\Psi} \;=\; 
{1\over {2 a}} \sum_{k=1,2}\; \{\;U^{k\ast}_{(i,j)-\hat{k}}
(D_k \Psi_{(i,j)-\hat{k}}) - (D_k \Psi_{(i,j)})\;\;\} 
\;+\;  {1\over 4} \kappa\sp2 
(\Psi_{i,j}\sp{\ast}  \Psi_{i,j} - 1)\Psi_{i,j}
\end{displaymath}
\begin{displaymath}
{1\over \beta } \dot {E^k_{i,j}} \;=\; 
{i\over {2 a}} [ U^k_{i,j}\Psi_{i,j+\hat{k}} \Psi_{i,j}\sp{\ast} - 
    U_{i,j}\sp{k \ast}\Psi_{i,j+\hat{k}}\sp{\ast}\Psi_{i,j} ]
 +  {1\over { a^3}}  \sum_{l \neq k} \{ sin [ a (
A^k_{i,j} + A^l_{i,j+\hat{k}} 
\end{displaymath}
\begin{equation}
 - A^k_{i,j+\hat{l}} - A^l_{i,j}) ]   - 
sin [ a (A^k_{i,j-\hat{l}}+
A^l_{i,j+\hat{k}-\hat{l}}-A^k_{i,j}-A^l_{i,j-\hat{l}}) ] \}
\label{latticeequations}
\end{equation} 
\noindent
Equations (\ref{latticeequations}) will be integrated numerically,
in order to follow the time evolution of any set
of initial data obeying the local constraint (\ref{latticegauss}).
Notice that no $ A_0 $ appears in the equations above.
They are written in the $ A_0 = 0 $ gauge,
which simplifies
considerably the equations of motion and 
makes the numerical task more tractable.
We leave for the Appendix 
the details about 
the lattice size, the time discretization and the 
integration algorithm used, 
and proceed with the presentation of our results. 

{\it (b) Vortex pair}

Let us consider two vortices with $N$ units of magnetic-flux each,
formed initially at a distance $d$ from one another. 
In terms of the fields $ \Psi $ and $ A_i $ it is
most convenient to take  
for the initial configuration the 
"product ansatz" of the corresponding two
axially symmetric vortex solutions \cite{rST}
\begin{equation}
\Psi ({\bf x}) \;=\;  \Psi\sp{(N)}(\vert {\bf x} - 
{{{\bf d}}\over 2}\vert)
\;  \Psi\sp{(N)}(\vert {\bf x} \;+\; 
{{\bf d}\over 2}\vert)   \nonumber \\
\end{equation} 
\begin{equation}
{\bf A}({\bf x}) \;=\;  {\bf A}\sp{(N)}(\vert {\bf x} - 
{{{\bf d}}\over 2}\vert)  \;+\;
{\bf A}\sp{(N)}(\vert {\bf x} \;+\; {{\bf d}\over 2}\vert) 
\label{vortexpairansatz} 
\end{equation} 
\noindent
The field $A_0$ is consistently set to zero, while the solution
of Gauss' constraint for the given $\Psi$ configuration provides us
with the initial data for the electric field 
$- \partial_t A_i ({\bf x}, t=0)$, 
necessary for the integration of the equations of motion.

Although for large separations this configuration 
is somewhat special,
being close to 
the minimum of the energy under the constraint 
of two zeroes in the scalar
field, for smaller $ d $ it imitates 
reasonably well the rather random
production of the vortices in a realistic situation.
(In fact, the axial symmetry of the 
"individual vortices" was relaxed in
several simulations. No deviations 
from the qualitative picture presented
below were observed).
Thus, the initial configuration consists 
essentially of two lumps of energy
and topological charge, concentrated 
around two local maxima at a distance
$d$ from one another. Strictly speaking, there is no unique
definition of a vortex position 
in a generic multi-vortex configuration.
The positions of the individual vortices are defined only
approximately, either as the positions of the zeroes of the 
scalar field, or as
the positions of the local maxima 
of the energy density or of the topological
density, or finally as the approximate guiding centers discussed
in the previous section.
All these are reasonable definitions
and their differences become less and 
less significant as one increases
the vortex separation. 
Having specified the initial configuration
one is ready to proceed with the numerical 
simulation of their motion.

We take for the parameters the values 
$\kappa=1.5$ and $\beta=0.04$ and
consider first the case of two minimal $N=1$ vortices.
Figures 1 and 2 show the results of
the simulation 
with the two vortices placed initially at
$(\,-\, 2.0\, ,\, 0\,)$ and $(\,+\, 2.0\, ,\, 0\,)$, respectively.
At this distance the vortices already interact significantly,
while retaining their individuality.
In Figure 1 the trajectory of the second vortex 
is plotted. 
The first one follows the image of the
above trajectory under reflection with respect to the origin.
To avoid overlapping of the 
trajectories, the run was interrupted after about 800 time units, 
when the vortices had each completed 
a full rotation around the origin. 
The picture that emerges is identical to the one obtained
in the study of the motion of a pair of electrons in the plane
and in the presence of a perpendicular 
magnetic field \cite{rS}, \cite{rF}.
Apart from an overall rotation
around the origin, a finer periodic 
motion can be perceived in the trajectory
of each vortex, shown in Figure 1. Borrowing the terminology
from the two-electron analogue, we give the name "cyclotron" to 
this finer motion of the vortices and we will study it in detail
later on.

The fact that the trajectories shown in Figure 1 are
so similar, indicates that the vortices
move like rigid bodies, without significant oscillatory
activity in their interiors. This is demonstrated beyond any
doubt also in Figure 2,
where we plot four snapshots of the energy density contours, 
corresponding to times $t=0$, $t=200$, $t=500$ and $t=700$, 
respectively.
In agreement with our previous conclusion, the shape 
of the energy profile in each of the two lumps 
seems to remain unchanged during the rotation.
A more detailed examination though, 
including animation of successive
snapshots, revealed a small 
oscillation in the sizes of the two energy lumps.
They spread a little and shrink periodically with a period equal 
to that of the cyclotron motion. 

An estimate of the accuracy of the results presented above
is obtained by examining the precision of the validity of 
the conservation laws 
during the evolution. Thus we consider next the time dependence
of the theoretically conserved quantities: energy, linear momentum,
angular momentum, and topological charge, as well as the local
constraint given by Gauss' law. 
After a complete period, the
total deviation in Gauss' law (the sum over all points of the grid 
of the absolute values of the 
local deviations) was less than $ 10^{-6}$.
>From the time evolution of the total energy $W$ of the system, 
depicted in Figure 3, one sees that it was conserved 
with an accuracy better than one part in $10^3$.
In contrast to the vividly oscillating four components 
$ W_e, W_b, W_d $ and $ W_v $,
also plotted in the same Figure, the total energy
is on the same scale a perfect 
straight line parallel to the time axis.
Ditto for the total angular momentum 
$l$ and its two gauge-invariant  
pieces $ l_1 $ and $ l_2 $, all plotted in Figure 4.
Although the two individual terms undergo rather wild 
oscillations, their sum is conserved to within a few percent.
The first term $l_1$ of the angular momentum (\ref{angularmomentum})
is $-\, \pi$ times 
the second moment of the topological density, a measure
of the size of the vortex pair. Thus, the oscillatory nature of
the cyclotron motion of the two vortices, shown in
Figure 1 above, is expected to induce
a similar behaviour in the time evolution of $l_1$, while
the conservation of the total angular momentum implies the
same for $l_2$. All with the same period.
This is trivially verified by a comparison of Figures 1 and 4.
Note that a periodical pattern
with the same period can also be detected in the energy plots after
careful examination. It seems that this oscillatory behaviour is a 
general feature of the system.
We turn next to
the total topological charge $ N $, whose time evolution 
is plotted in Figure 5.  $ N $ starts at $t=0$ with the 
value $ 1.993 $, and up to a small fluctuation of
less than one part in $10^3$ it retains that value all
during the numerical simulation. 
Its deviation from 
the continuum value
$N=2$ is due to the spatial discretization of the system.
Finally, the position of the guiding center of the system (or 
equivalently its total linear momentum) is considered. This 
is a conserved quantity, which, due to the 
symmetry of the starting configuration,
initially coincides with the origin of the coordinate 
system.
It was checked to be pinned there with impressive accuracy 
all during the simulation.
 
Thus, the picture that arises clearly confirms the 
theoretical predictions discussed in the previous section.
All the conserved  quantities of the continuum  are
respected with high accuracy. Furthermore the system is 
characterized 
by periodic patterns manifested
in the trajectory plots, one example of which is what
we called "cyclotron motion".
This whole qualitative picture is generic. 
It was verified in all the simulations performed, 
for a large variety of initial configurations 
and for a wide range of values of the parameters.

We proceed next to the quantitative comparison 
of our numerical results
with the rough theoretical prediction (\ref{Omega}) of the 
angular velocity of the vortex rotation. For that one needs the
interaction potential $U_{\rm vv}(d)$ between the two vortices 
as a function of their distance. Define $U_{\rm vv}(d)\;=\;E(d) - 
2 E_{N=1}$, with $E(d)$ the minimum of the
energy in the $N=2$ sector with the constraint that the scalar 
field vanishes at two points, a distance $d$ apart, 
and $E_{N=1}$ the energy of the single vortex solution.
The result for $\kappa = 1.5$ and $\beta = 0.04$ 
is plotted in Figure 6. 
One sees that for the values of the parameters chosen above, 
$ U_{\rm vv}(d) $ is repulsive at all distances, falling to zero
very quickly.
With the interaction energy $ U_{\rm vv}(d) $ 
at hand one may calculate 
numerically its derivative and plot the theoretical
prediction for the period of revolution derived 
from the right hand side
of (\ref{Omega}). This is illustrated 
by the continuous curve in Figure 7.
One then simulates numerically the motion of the vortex pair 
for various initial separations and from the time it takes for
them to cover a full circle around each other one determines 
the corresponding period. The result is represented by the
little triangles on Figure 7.  
The agreement is quite remarkable, 
down to distances of the order of
the vortex characteristic diameter, at which the two vortices 
overlap almost to the point of loosing their individuality.

It was pointed out in reference \cite{rST}, 
that in contrast to the vortex-
antivortex which always attract each other \cite{rC},
the interaction energy $U_{\rm vv}(d)$
between two vortices is not in general 
a monotonically decreasing 
function of their distance. For instance, 
the potential $U_{\rm vv}(d)$ 
for the model with $\kappa=0.5$ and $\beta=0.005$ shown in Figure 8,
increases up to a local maximum at $d \simeq 7$ and decreases 
beyond that.
In agreement with (\ref{Omega}) one then expects the two $N=1$
vortices to rotate counterclockwise 
when put at a distance greater than
7, and clockwise when the initial separation is smaller than 7.
This is exactly what is observed for two vortices placed initially
on the x-axis, symmetrically with 
respect to the origin at a distance
$d=8$ and $d=4.5$, respectively. 
The trajectories of the vortices initially 
on the right, as determined
by the zero of the scalar field, for both
simulations are shown in Figure 9. The absolute values of the
corresponding angular velocities 
are also in agreement with (\ref{Omega}).  
The rather vivid fluctuations in the details of the two interacting 
vortices is a general
feature in small $\kappa$ models.

Before we move on to the discussion of the vortex-antivortex
system, and in order to improve one's 
intuition about the behaviour of the 
vortex-pair,
we would like to push a bit further its qualitative 
analogy with the two-electron system. For that we will study
and compare the details of their cyclotron motions.
The simplicity of the 
electron system allows for a complete analytical treatment and
for a detailed description of their trajectories.
One finds \cite{rS}, \cite{rF} that 
generically the corresponding guiding 
centers perform circular motion, while
the orbits of the electrons themselves exhibit 
patterns similar to those of Figure 1. Furthermore, it can be shown
that for given initial conditions, 
the characteristic wavelength 
and width of their cycloid motions increase 
when one 
decreases the strength of the external 
magnetic field, or as one increases
their mutual interaction by decreasing their
separation, or by 
increasing their electric charge.
A very similar picture emerges in the vortex-pair case. To study it
we first performed a series of simulations for various $d$'s, 
for the same values of the
parameters $\kappa=1.5$ and $\beta=0.04$.
The results depicted in Figure 10, combined with Figure 6, 
clearly confirm the claim that the
cyclotron wavelength and width both 
decrease with the vortex-vortex force.
Next, we varied $\beta$ and followed 
the trajectory of the energy maximum
corresponding to one of the two vortices for the same time 
interval in
all cases. From the results of the simulations shown in Figure 11,
it becomes apparent that an increase of 
$\beta$ leads to a 
decrease of the corresponding cyclotron characteristics. 
Notice also
from the same figure and equation $\Omega$, 
that the intervortex force increases with $\beta$.
We do not show it here but by experimenting with other values
of the vortex flux $N$ one may verify in the details of the
cyclotron motion, the correspondence mentioned above of
$N$ to the external magnetic field $B$ of the two-electron 
planar system.

\newpage
{\it (c)  Vortex-antivortex system}

We next replace one of the vortices
of the pair by an $N=-1$ antivortex and let it evolve. 
Like in the analogue electron-positron planar system 
with the external
magnetic field and for reasonable initial velocities, 
we expect that, again up to a small cyclotron
motion, the vortex and the antivortex will move
in formation along parallel trajectories. This picture 
contradicts naive
intuition, according to which the vortex and the antivortex would
as a result of their attraction approach each other and
gradually annihilate into elementary excitations.

A series of simulations was performed with the vortex-antivortex
system.  The initial field configuration used, was the product
superposition of an 
$N\,=\,1$, and an $N\,=\,-1$ axially symmetric vortex solutions,
according 
to the ansatz
\begin{eqnarray}    
 &    \Psi ({\bf x}) \;=\;  \Psi\sp{(1)}(\vert {\bf x} - 
	{{{\bf d}}\over 2}\vert)
\;  \Psi\sp{(-1)}(\vert {\bf x} \;+\; {{\bf d}\over 2}\vert)  
& \nonumber \\
 &    {\bf A}({\bf x}) \;=\;  {\bf A}\sp{(1)}(\vert {\bf x} - 
	{{{\bf d}}\over 2}\vert)  \;+\;
   {\bf A}\sp{(-1)}(\vert {\bf x} \;+\; {{\bf d}\over 2}\vert)  & 
\label{vortexantivortexansatz}
\end{eqnarray} 
\noindent
where $ \Psi\sp{(1)}, {\bf A}\sp{(1)} $ are the static 
fields of the 
$N=1$ vortex and $ \Psi\sp{(-1)}, {\bf A}\sp{(-1)} $ 
are those of the $N=-1$ antivortex. 
Like in the vortex-vortex case, $A_0$ is set to zero and
the initial values of the electric field are obtained by solving
Gauss' equation with the above $\Psi$.

The trajectories of the vortex and the antivortex in a typical run 
are illustrated in Figure 12. Both the approximate guiding centers
and the local maxima of the energy density were followed
and are shown on the same plot. 
The parameters were set to $\kappa \,=\, 1.5$ and 
$ \beta \,=\, 0.04$,
while the initial positions of the vortex and the 
antivortex (zeroes of
the scalar field) were taken
at $ (-2,10) $ 
and $ (2,10) $, respectively.
The vortices moved in the negative y-direction for 
$20$ space units,
i.e. about sixteen times their size, while retaining 
their initial shape and 
keeping their initial separation constant.
They moved with constant speed 
$ V \,=\, 0.025 $.  
The vortex velocity in the 
vortex-vortex pair for the same values of parameters and 
separation, was found to be $ V \,=\, 0.016 $. The higher velocity
in the vortex-antivortex case implies a potential between them
steeper in absolute value than the one in Figure 6.

Apart from the parallel transport of vortices, 
one sees a finer oscillating
motion, the "cyclotron motion" we mentioned in the preceding
paragraphs. We ran several simulations to study its details,
for various values of  $d$ and $\beta$. 
As in the vortex pair system,
both wavelength and width of the oscillation 
are decreasing functions of
$d$ and $\beta$, in perfect qualitative 
agreement with the dependence
of the cycloid patterns in 
the electron-positron analogue. 

We would like to end the discussion of the results
with a final comment
about boundary effects. As it follows from Figure 12, 
if we neglect the 
cyclotron motion, the paths of the two 
vortices are perfect straight lines parallel to the y-axis. 
A slight convergence of the trajectories 
towards each other appears though,
when the vortices come close to the boundaries.
Actually, if they start $6-7$ space units away from the boundaries, 
the initial convergence disappears,
to appear again when the vortices get 
close to the negative-y boundary.
By trying different sizes of grids, 
one concludes that this behaviour 
is a boundary effect sensitive to the absolute 
separation between the
vortices and the boundaries,  but insensitive 
to the size of the grid.
In the case of relatively small grids, the convergence of the paths
could be misleadingly interpreted as a generic feature 
of vortex dynamics.
To avoid this effect in our simulations, we always placed 
the vortices sufficiently away from the boundaries of the grid.

\section{Discussion}

The direct numerical simulation of the motion of a
generic vortex-(anti)vortex configuration
confirms the Hall behaviour, predicted analytically in a
previous publication.
The quantitative agreement persists even when
the two solitons overlap to the point that they can hardly be
considered as two.
Physically, this behaviour may not be entirely surprising.
It might be described as the well known Hall effect. After all,
the vortex of the model is microscopically \cite{rST} 
a non-vanishing electric charge density,
which is sustained by the non-linear forces 
(attractive electrostatic
and $\Psi$ self-interactions), and
circulates around its center, thus giving rise to the vortex 
magnetic field.
The current is locally perpendicular to the 
electric field and hence
consistent with the absence of energy dissipation.
Thus, the overall situation looks similar to the
ordinary Hall setting, only the circulating 
charges are immersed in their
own magnetic field and repelled by it, instead 
of being kept in orbit
by an externally prescribed one and quite naturally, 
a vortex is expected to exhibit the Hall behaviour described here. 
One may push the picture even farther by noticing that since 
the charges as described by the 
wave-function $\Psi$, are spread over the entire 
region of the vortex,
they feel the integral of the magnetic field, i.e. 
the winding number
of the configuration, which makes plausible the appearance of $N$
in formulas  (\ref{momentumalgebra}) and (\ref{vortexvelocity}). 

Mathematically on the other hand, one is dealing with the 
most general model describing the dynamics of a 
condensate wave-function $\Psi$ coupled to
the electromagnetic potential,
and restricted only by the translational, rotational
and gauge invariance of the system. The ion lattice
assumed frozen, defines
a preferred reference frame and breaks the 
Poincar$\acute {\rm e}$
invariance of the underlying fundamental system.
Topological or metastable non-topological solitons \cite{rBTb} in
models with just these symmetries \cite{rLee} 
are expected \cite{rST} to exhibit identical
Hall behaviour. This is indeed what happens in 
all the systems examined 
so far
\cite{rWZ}, \cite{rDI},
even in ferromagnets which have no physical similarity to a 
system of charges interacting 
with the electromagnetic field \cite{rPTa}, \cite{rPZ}.

Clearly, the next step is to test the predictions 
of the model at hand against more realistic experimental situations.
One should study the static properties of vortices in thin films 
with finite thickness, and then analyse 
their response to an external
current in the context possibly of an improved model to incorporate
dissipation.

\newpage

{\bf Appendix: The Numerical Algorithm }

To solve the initial value problem defined by the system of 
equations (\ref{latticeequations}) and a starting 
configuration of the form (\ref{vortexpairansatz}) or 
(\ref{vortexantivortexansatz}) we considered in this paper,
we used a 
leapfrog updating scheme \cite{NumericalRecipes}
where the time levels in the time derivative term 
"leapfrog" over the
time levels in the space derivative term. Equations 
(\ref{latticeequations}) is a 
mixed system of first order and second 
order differential equations in time. 
A leapfrog algorithm for a second order 
equation is equivalent to the
updating of fields and momenta successively, 
but the coupling of that  
equation to a first order equation demands special 
care in the construction 
of the algorithm. Nevertheless the leapfrog algorithm gives marked
improvement in stability over the simpler approach of updating 
both fields and their momenta at the same time level. 

To perform our simulations  we found a $161 \times 161$ grid, with
lattice spacing $a \;=\; 0.15$ of sufficient accuracy. 
The space resolution of that grid is estimated by calculating 
numerically the total topological charge $N$ of 
(\ref{vortexpairansatz})
and comparing it to
the exact value $N  \,=\, 2 $. 
Using formula (\ref{topologicaldensity}) for the topological charge
one finds $N\,=\,1.993$.
Interestingly, the alternative formula for 
$N\,=\,(1/{2 \pi}) \int d^2x B$
is less sensitive to the discretization and gives
$N\,=\,1.999$ for the initial configuration.
The accuracy of the simulations
is further estimated by the conservation 
of energy which is respected
over one period's time with accuracy better 
than $ 0.1\% $ in all our 
runs.  We imposed Neumann boundary 
conditions by  setting,  the 
covariant derivative in the normal 
to the boundary direction, equal 
to zero. To do so we fixed the value of the Higgs 
field at each point
of the external layer of the grid equal to 
their first inner neighbour.
Also the values of the gauge fields, which live at the links 
which connect those neighbours, were set equal to zero. 
To test our results
we ran simulations in bigger grids $ 251 \times 251 $ 
with the same
or smaller lattice  spacing, say $a\,=\,0.1$, 
and the results obtained
were all perfectly consistent. We used a bigger 
grid $ 251 \times 251 $ and $a \;=\; 0.15$ in the 
vortex-antivortex  simulations in order to follow the orbits of 
the vortices for longer distances. The time step $\Delta t$ we used 
in most of our runs was 0.001 or 0.002 but the algorithm was stable 
and accurate for even bigger time steps. All our simulations were 
performed on various HP workstations in Crete. 
A typical run of duration 
$T \approx 800$ time units, with  $\Delta t\,=\, 0.002 $
on a $161 \times 161$ grid, needed about 80 hours of CPU time 
on a HP-735 machine.
    
Finally we wish to comment on our choice to use the formalism
developed in the study of lattice gauge theories.
One may envisage two discretization schemes to convert the equations
of motion of our theory into difference
equations. The conventional discretization
scheme (CDS) and the lattice gauge formalism (LGF). We experimented
with both and finally adopted the latter for its elegance and
functionality.
It should be pointed out that both methods 
have been used in the study of vortex dynamics in relativistic
models \cite{rShel}, \cite{rMatzner}, 
\cite{rMCarlo}, \cite{rNCode}, \cite{rMRS} with satisfactory and
consistent results. The LGF, especially designed to preserve
the local constraint, is certainly more natural to use in a gauge
theory, but for our problem there was another more serious issue to 
face. Use of any CDS explicitly violates gauge invariance. 
Without the gauge invariance there is no reason 
for the local constraint to be satisfied. In fact, 
a violation of the equation of continuity and of Gauss' 
law was obtained,
which in addition was accumulative in our simulations based on any
CDS we tried.
Whenever the error in those became significant, the integration
routine destabilized. 
The way out in the context of a CDS would be to
use a sufficiently small time grid spacing to retain the error at
sufficiently small values all during the time interval
required for the study of the phenomenon of interest. 
This has worked reasonably well in the
study of vortex scattering in
relativistic models \cite{rShel}, \cite{rMatzner}, 
since the process takes very little time
and one is able to see the phenomenon on the computer without
an excessive consumption of CPU time. 
In our case though this did not work. 
The vortices rotate very slowly
around each other and in order to see a full turn one has to
wait for a long time. In fact for a much too long time for
any CDS we tried to be stable.

{\bf Acknowledgments}

We would like to thank Professor N. Papanicolaou for
several helpful discussions. G.N.S would
like to acknowledge the hospitality of the Department 
of Applied Mathematics of University of Durham, 
where part of this work was performed.  This research was supported 
in part by the EU grants 
CHRX-CT94-0621 and CHRX-CT93-0340, by 
the Greek General Secretariat
of Research and Technology grant No $91 \Pi {\rm ENE}\Delta 358$
and by a research fund from the University of Crete.

\vfill

\newpage

\vskip 2cm
\hskip 0.5cm {FIGURE CAPTIONS}

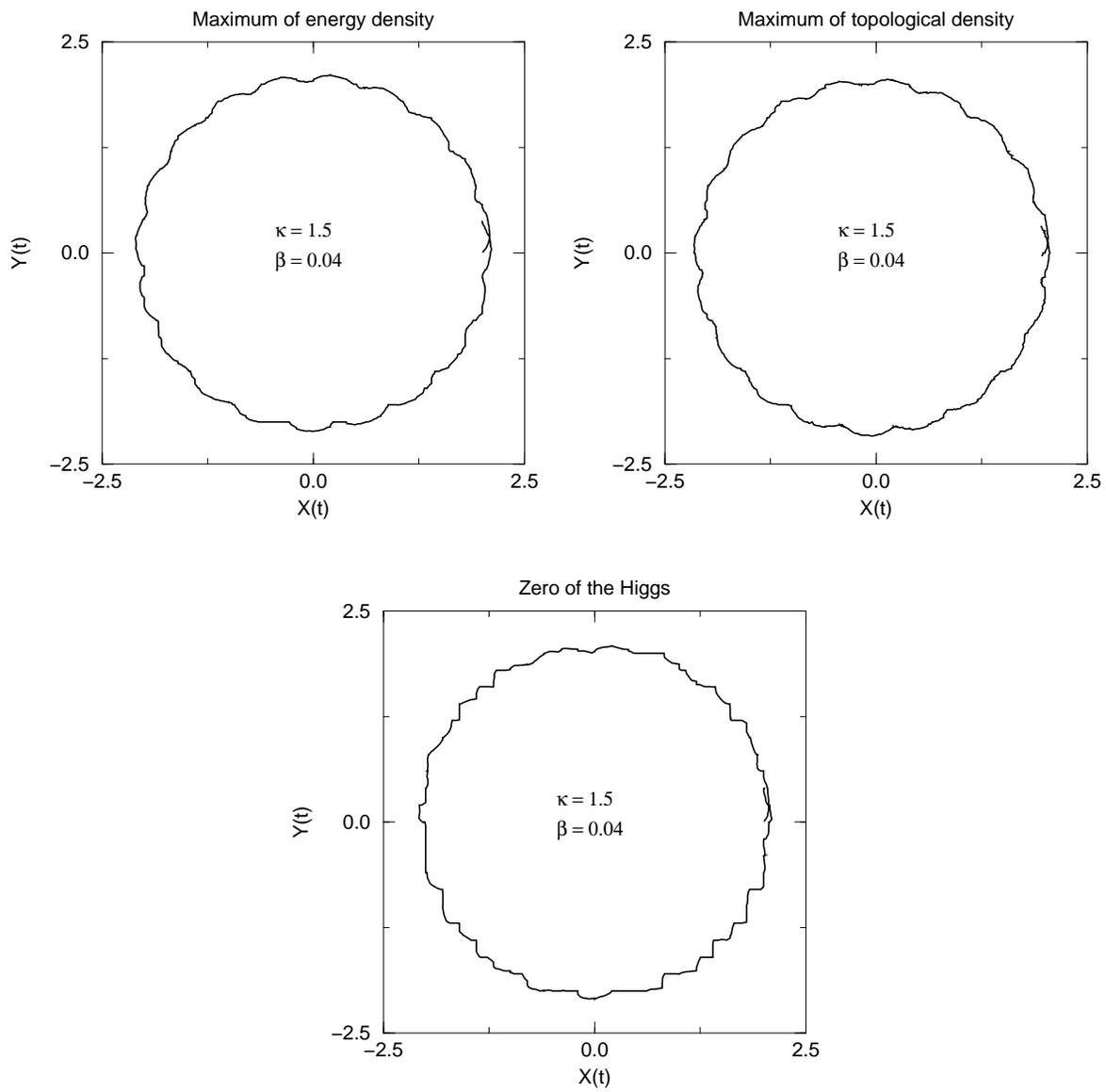
Figure 1: The trajectory of one vortex in the pair as 
determined by the
location of the maximum of the energy density, 
of the maximum of the topological density and of the zero of the
scalar field $\Psi$. 
Time duration t=800.

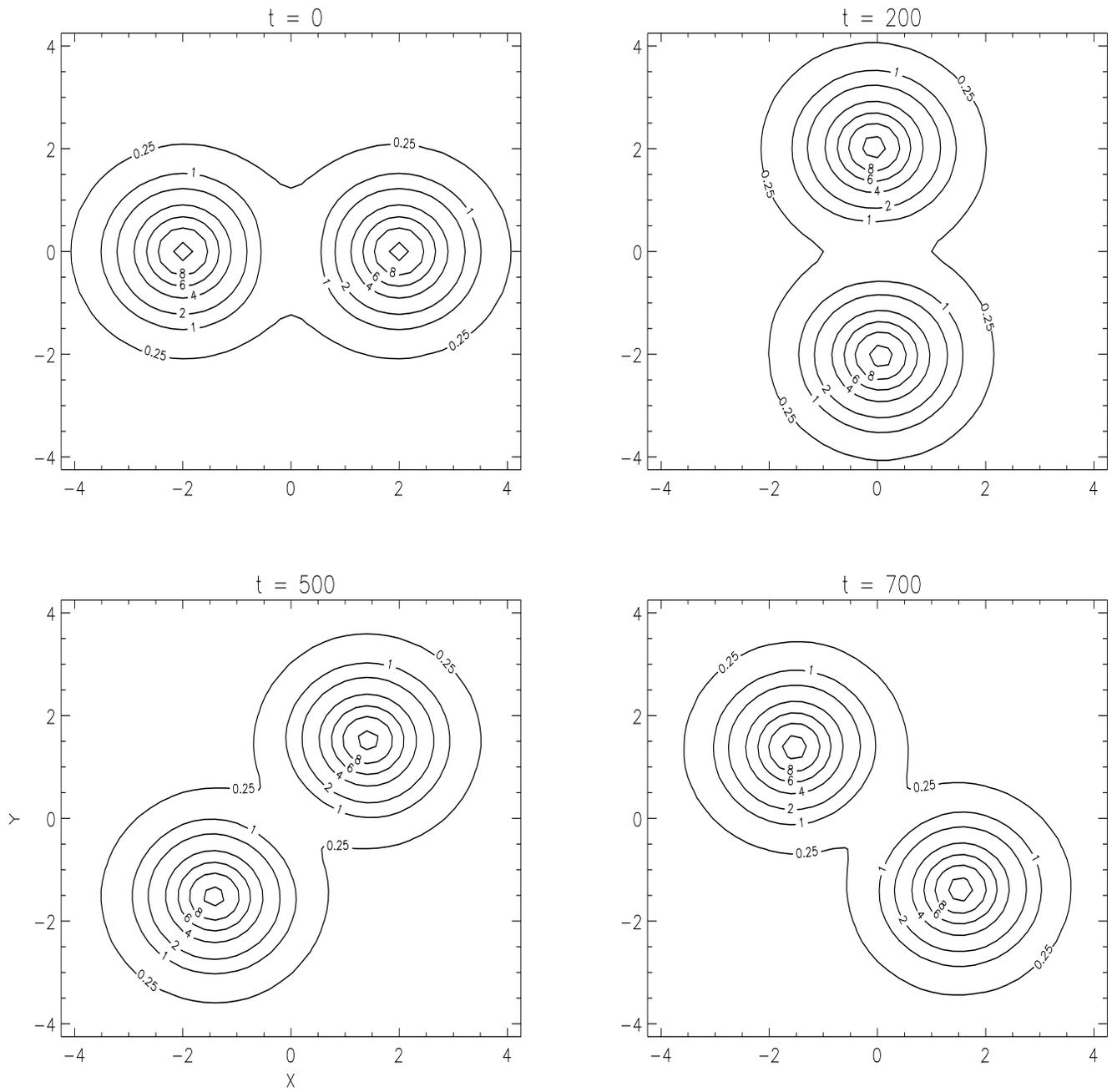
Figure 2: Four snapshots of the energy density contours 
during the motion of the vortex pair.

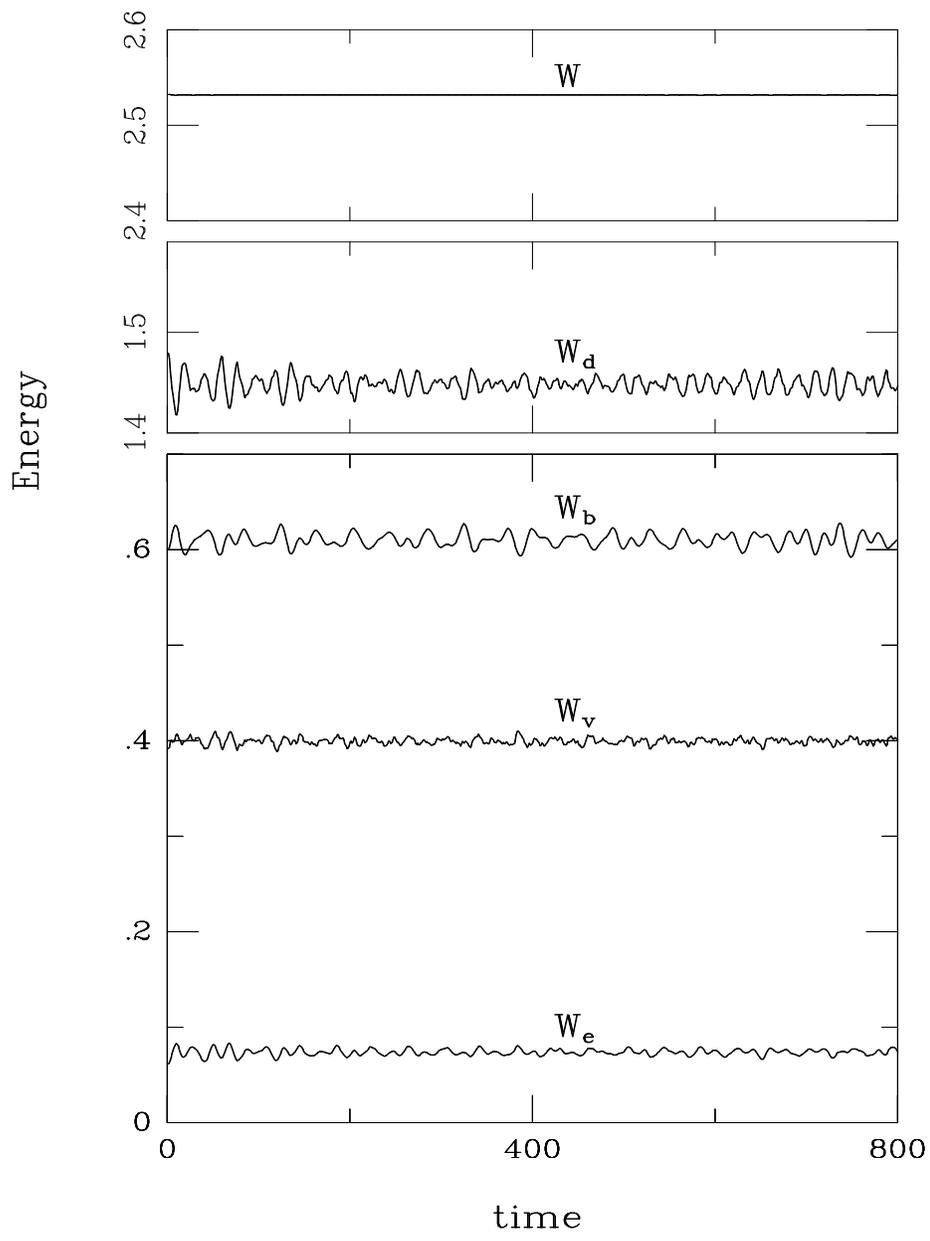
Figure 3: The time dependence of the various 
components of the energy
of the vortex pair and of the total energy W. 
Note the high accuracy in the
conservation of the total energy of the system.

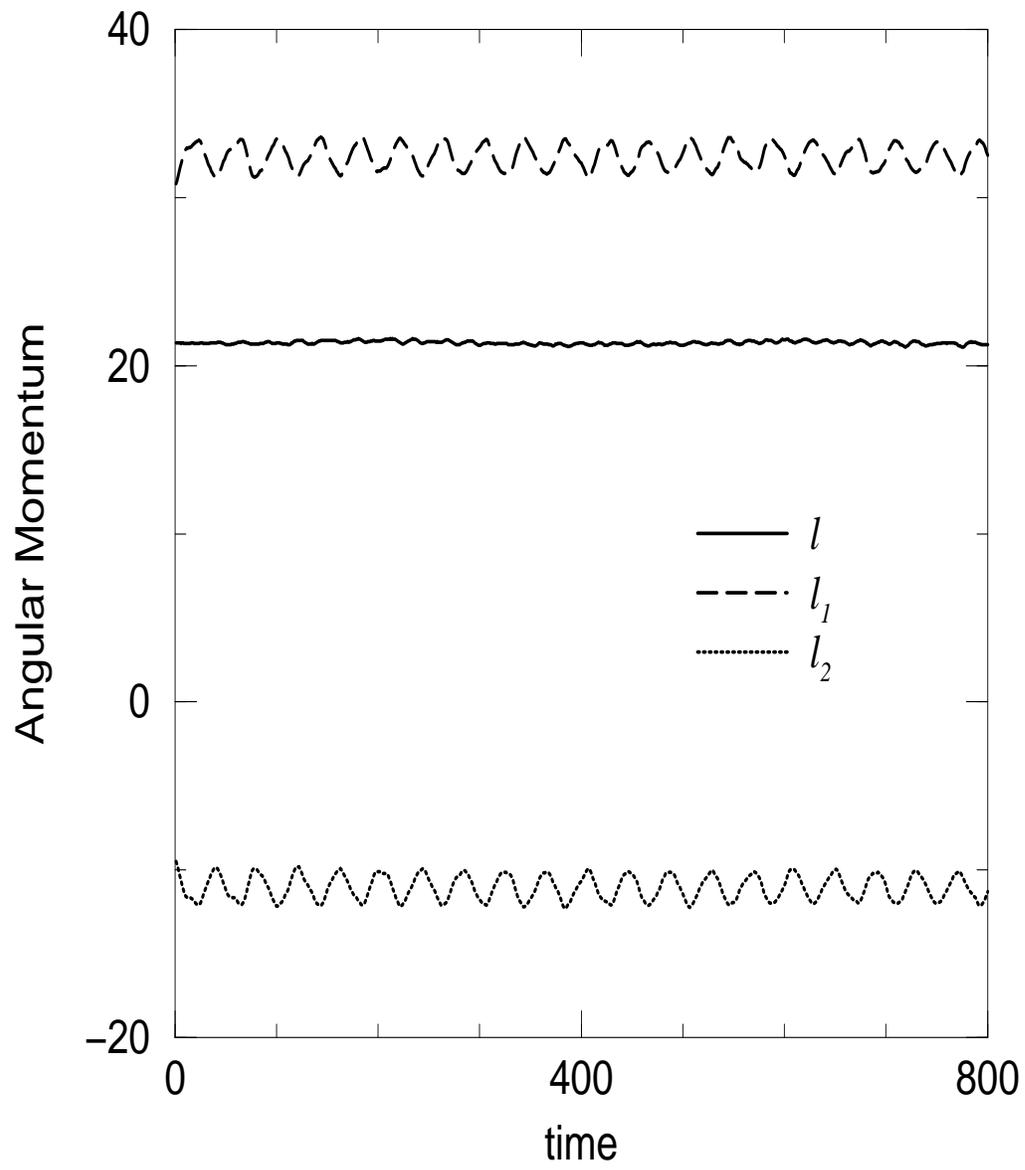
Figure 4: The two pieces of the angular momentum and their sum. The
conservation of the total angular momentum to 
within $2 \%$ is quite satisfactory.

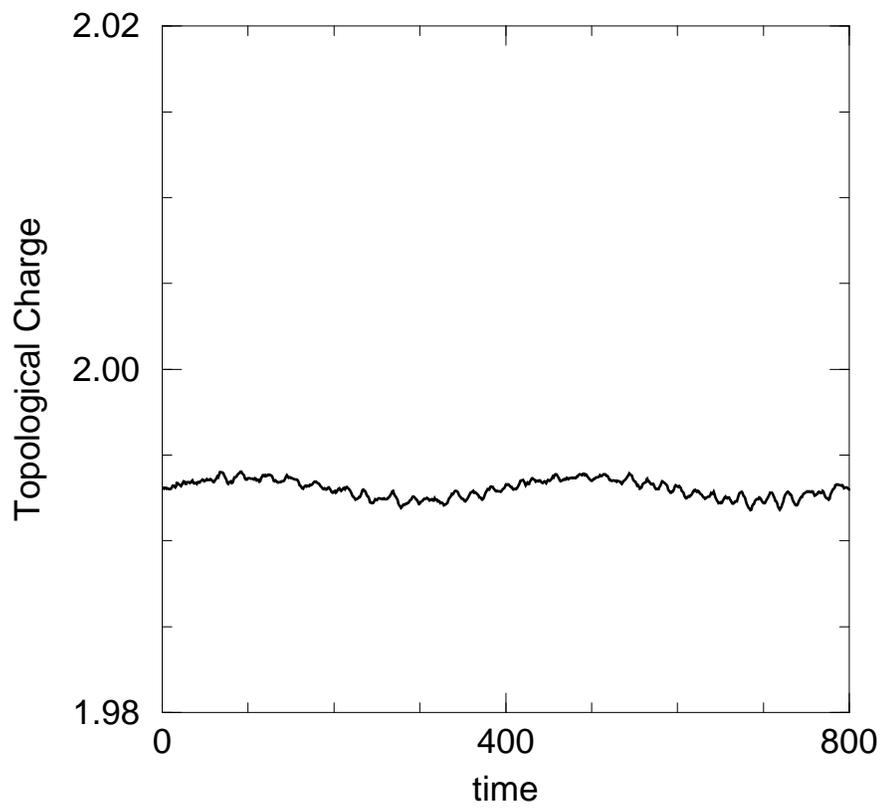
Figure 5: The time evolution of the total topological charge 
of the vortex pair.

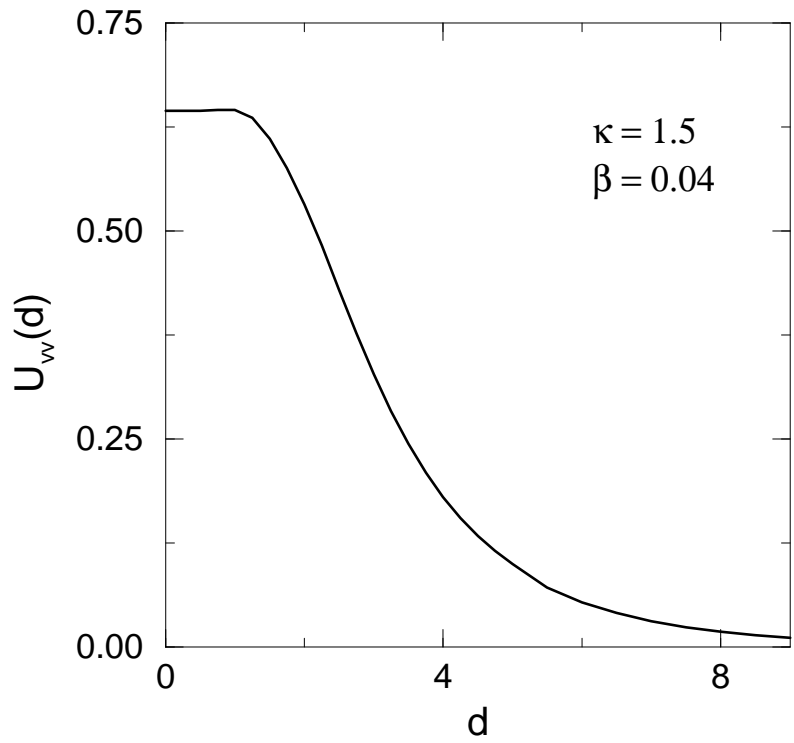
Figure 6: The interaction energy of two 
$N\,=\,1$ vortices for $\kappa=1.5$ and 
$\beta=0.04$ as a function of their separation $d$.

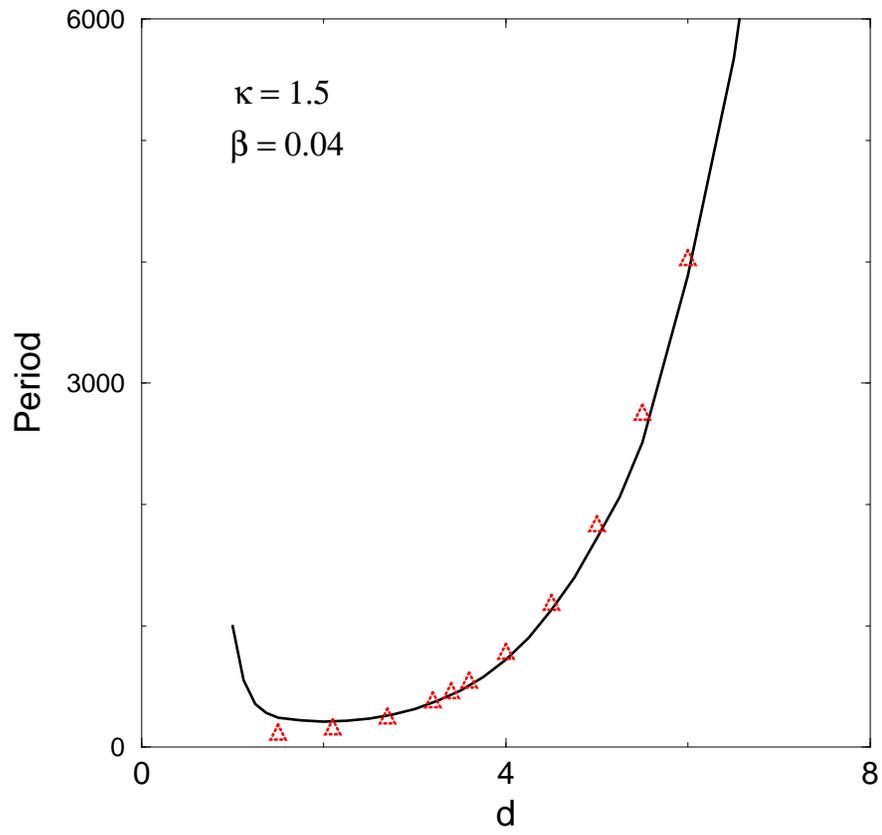
Figure 7: The period of revolution of the vortices 
around each other
as a function of their separation, 
computed from the theoretical formula and 
the slope of the curve
of Figure 6 (solid line) and from the numerical 
simulation (triangles).

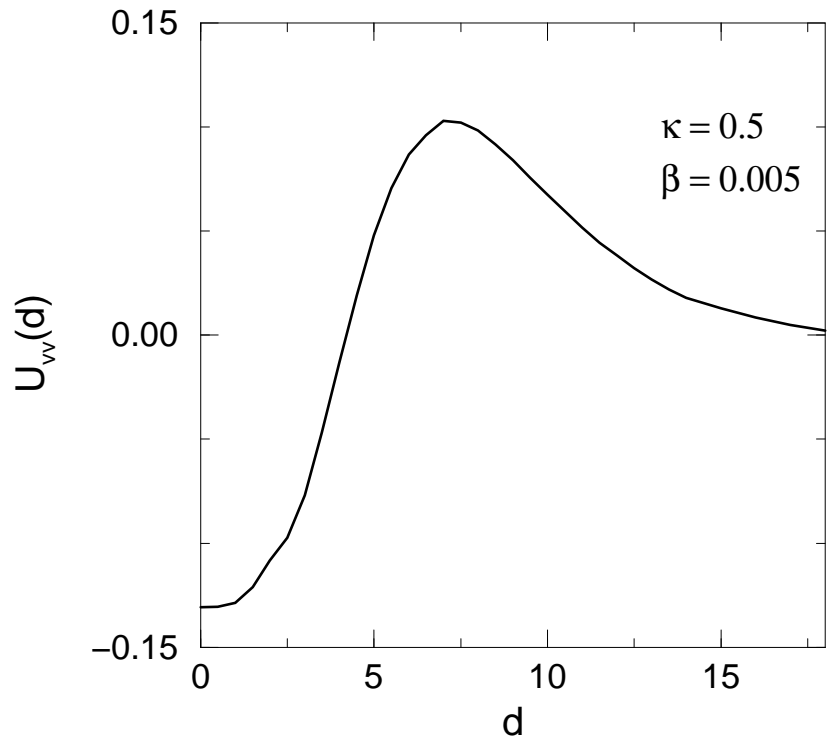
Figure 8: The interaction potential of the two 
vortices for $\kappa=0.5$
and $\beta=0.005$. The potential is attractive 
for small distances and
repulsive for $d\,>\,7$.

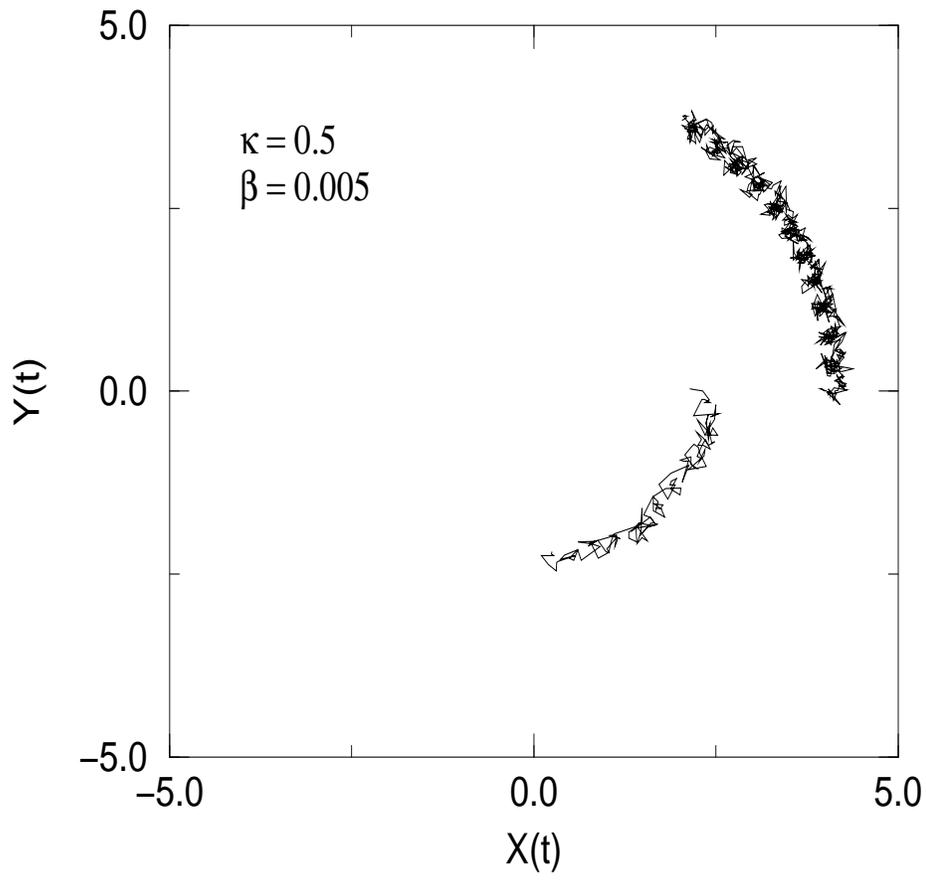
Figure 9: In agreement to the theoretical prediction, 
when the $N\,=\,1$ vortices attract each other, they rotate 
clockwise, while when they repel, they 
rotate counterclockwise.

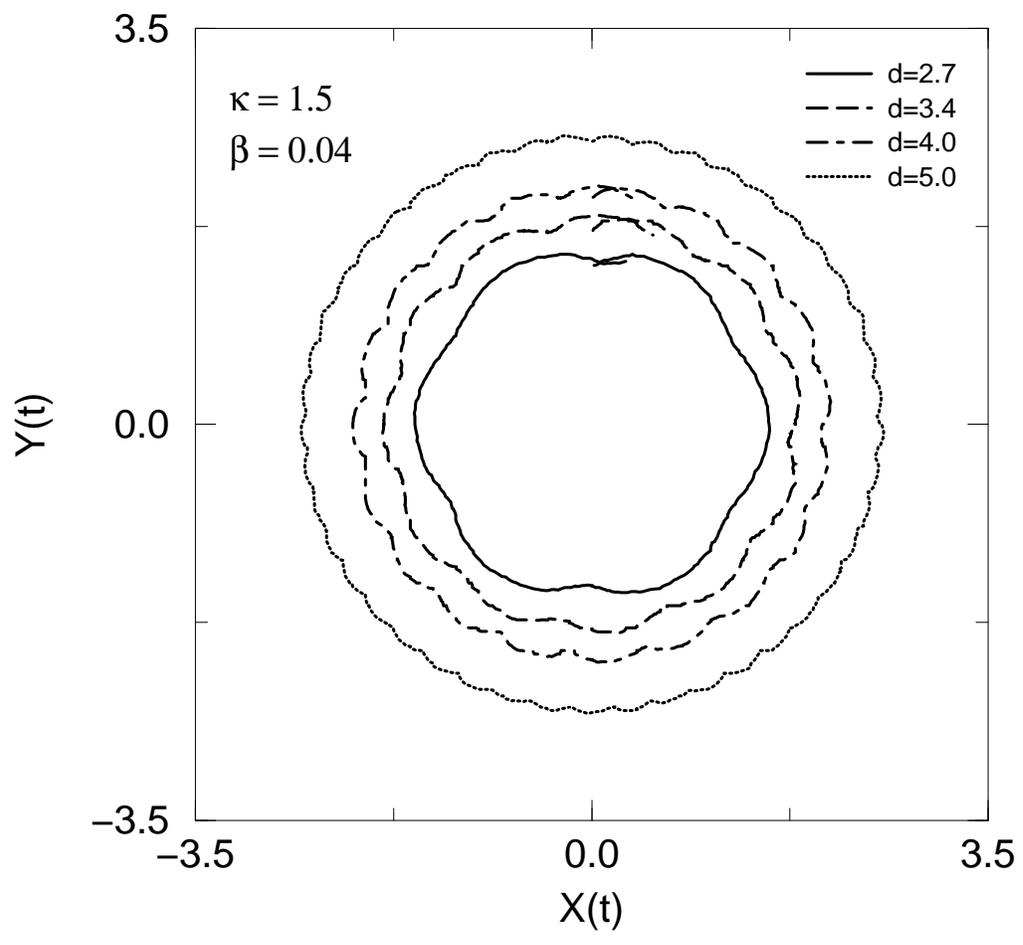
Figure 10: The trajectory of the maximum 
of the energy density of one
of the two vortices of the pair. The  
dependence of the characteristic wavelength and width 
of the cyclotron motion of each vortex
on their separation d is qualitatively identical to 
the one obtained in the two-electron system.

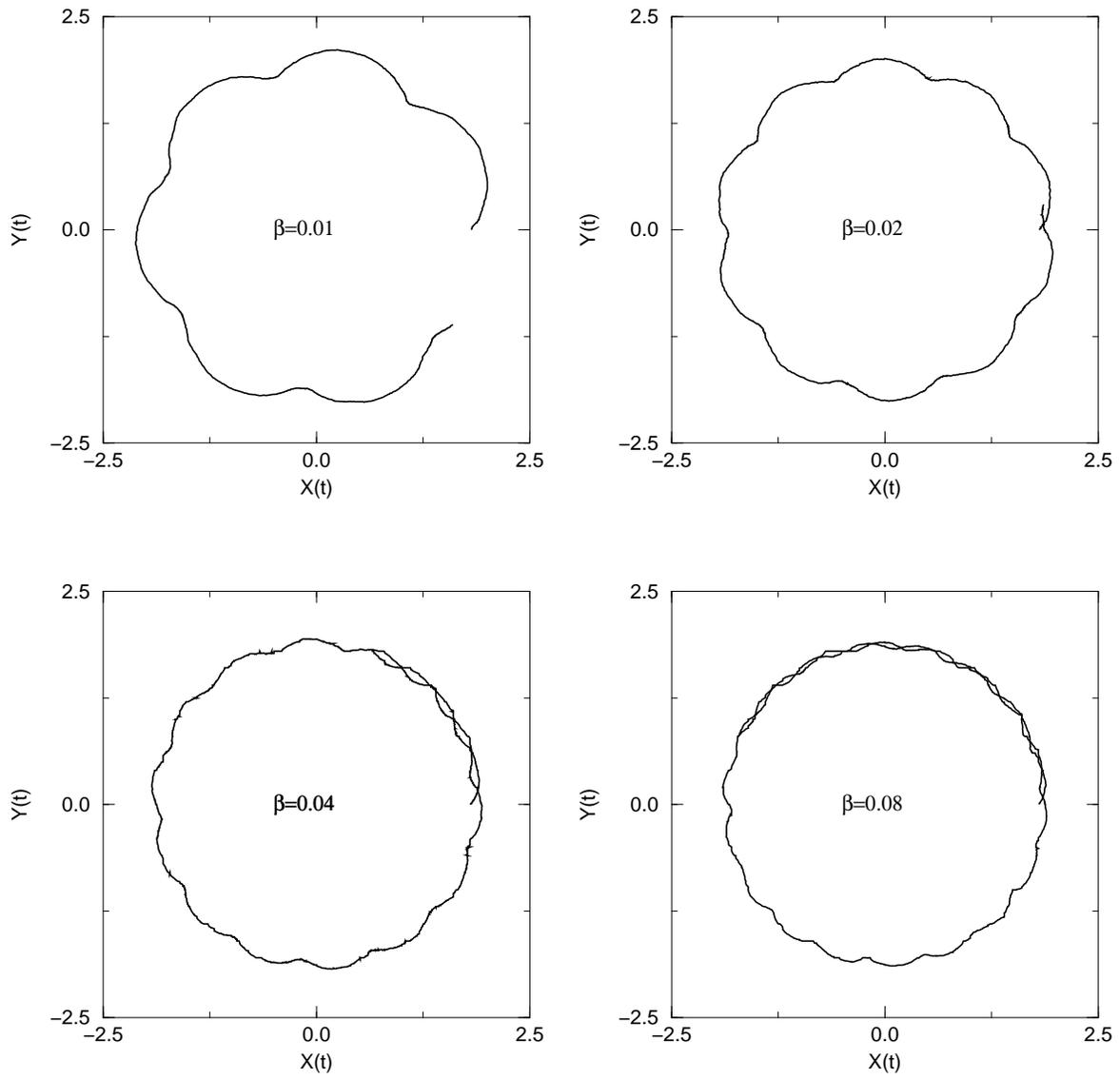
Figure 11: The trajectory of the vortex, as it is determined from
the position of the corresponding energy maximum, 
for various values of the parameter $\beta$ and for the same total
duration in all runs.

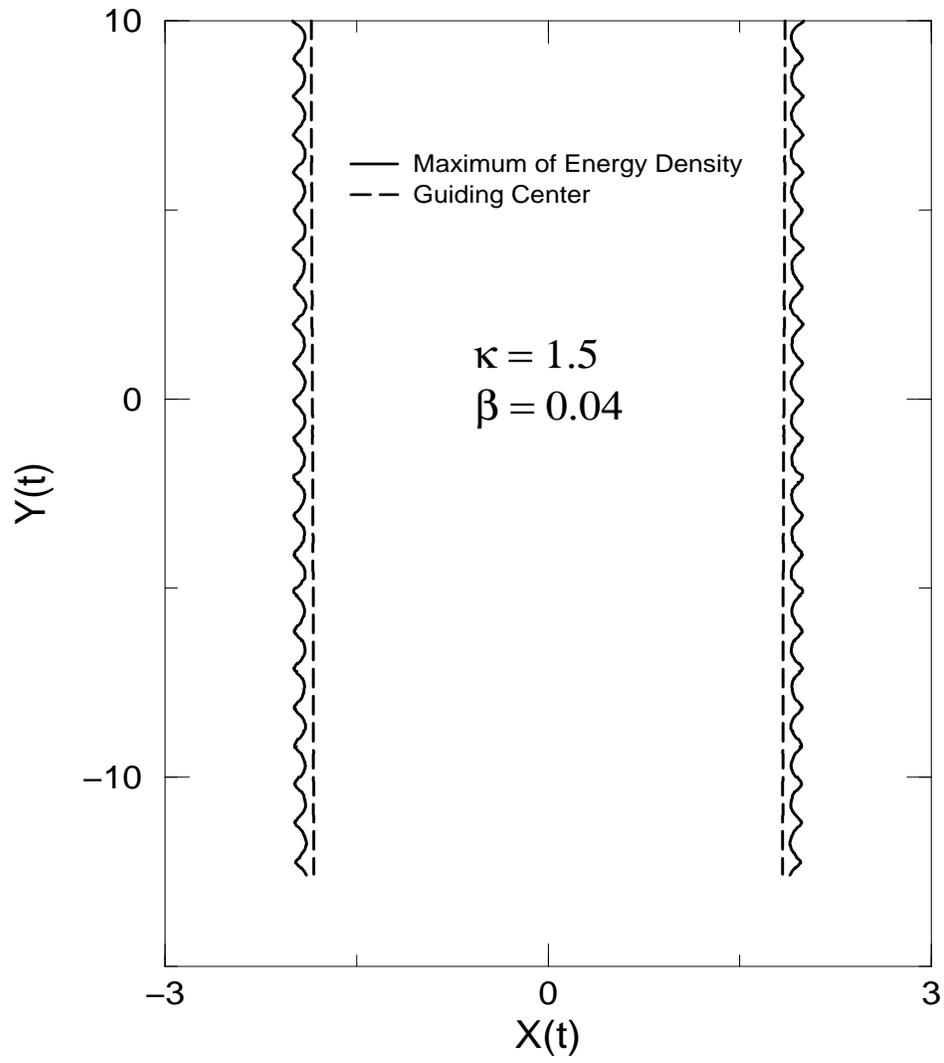
Figure 12: The evolution of a typical initial vortex-antivortex
configuration. The wavy and the straight lines are the trajectories 
of the energy density maxima and of the 
approximate guiding centers, respectively.

\end{document}